	\documentstyle[twoside,epsfig]{hsproc}

\newcommand{\beq}{\begin{eqnarray}}
\newcommand{\eeq}{\end{eqnarray}}
\newcommand{\no}{\nonumber}
\newcommand{\D}{{\mathcal D}}

\newcommand{\vx}{v\!\!\:\cdot\!\!\: x}

\newcommand{\vp}{v\!\!\:\cdot\!\!\: p}
\newcommand{\vd}{v\!\!\:\cdot\!\!\:\partial}
\newcommand{\vD}{v\!\!\:\cdot\!\!\:{\mathcal D}}

\newcommand{\V}{{\mathcal V}^\pi}
\newcommand{\A}{{\mathcal A}^\pi}

\def\slash#1{#1 \hskip -0.5em / }

\def\Pp{\frac{1 + \slash{v}}{2}}

\def\gl#1{(\ref{#1})}

\def\L#1{{\cal L}_{\mbox{\scriptsize #1}}}

\def\tr#1{{\rm tr}\left[#1\right]}

\begin{document}

	\title{Heavy Hadron Spectrum and Interactions}
	\author{D.\ Ebert\footnote{Supported by {\it Deutsche 
	Forschungsgemeinschaft} under contract Eb 139/1-2.}
	 and T.\ Feldmann\footnotemark[1]}{Institut f\"ur Physik, 
	Humboldt-Universit\"at zu Berlin,\\ Invalidenstra\ss{}e 110,
	D-10115 Berlin, Germany}
%

\vspace{3em}
\begin{center}
\noindent
Talk presented at \\
{\it 
International Conference on Hadron Structure:\\
 High-Energy
Interactions: Theory and Experiment (Hadron Structure 96),\\
 Stara Lesna, Slovakia, 12-17 February 1996. \\
(to appear in the proceedings) }
\end{center}
\vspace{3em}

	\abstract{Starting from the approximate symmetries of QCD,
	namely chiral symmetry for light quarks and spin and
	flavor symmetry for heavy quarks, we investigate
	the low-energy properties of heavy hadrons.
	For this purpose we construct a consistent picture
	of quark-antiquark and quark-diquark interactions as
	a low-energy approximation to the flavor dynamics
	in heavy mesons and heavy baryons, respectively.
	Using standard functional integration tools, we
 	derive an effective Lagrangian in terms of heavy
	hadron fields and discuss several properties, like
	the mass spectrum, coupling and decay constants,
	Isgur-Wise form factors.}
%
	\def\runauthor{D.\ Ebert, T.\ Feldmann}
	\def\shorttitle{Heavy Hadron Spectrum and Interactions}
%
%
%

\section{Motivation}

In the framework of the Standard Model,
the theory of strong interaction is given by
Quantum Chromodynamics (QCD), which is
formulated as a renormalizable
non-abelian gauge theory in 
terms of quark and gluon fields.
Due to asymptotic freedom of the theory, the
short distance part of the QCD dynamics can
be treated perturbatively.
However from the phenomenological point of view,
one would like to formulate the theory
in terms of the one-particle states
entering the detectors in high-energy experiments,
which are the hadrons rather than quarks and gluons.
The long-distance QCD effects leading to the confinement
of quark and gluons and their binding to hadrons,
can only be understood non-perturbatively, and
therefore it is a big challenge to derive effective hadron
lagrangians directly from QCD.
Such a program evidently requires
(possibly crude) approximations on the long way from QCD down to the
effective hadron theory. Simplifications arise, however, since we are
naturally restricted to the low-energy region.
Therefore, from the practical point of view, it is sufficient to
find an approximation to QCD which mimics the
essential features of low-energy quark flavor dynamics. 
Clearly, the possible form of such effective quark
lagrangians must be restricted by the underlying
symmetries of QCD, which should be viewed as a guide
to find tractable models of quark flavor dynamics.

In the sector of light quark flavors
$q=(u,d,s)$, QCD possesses an approximate $SU(3)_L \times
SU(3)_R$ chiral symmetry which is spontaneously broken
to $SU(3)_V$, leading to the emergence of (pseu\-do)\-Gold\-stone
bosons $\pi,K,\eta$, which receive their masses by the explicit
breaking of chiral symmetry through current quark masses.

Recently, new important symmetries have
also been discovered for heavy quark flavors $Q=b,c,\ldots$
which considerably simplify the description of 
hadrons containing one heavy quark.
These symmetries arise
in the limit of infinite heavy quark masses $m_Q \rightarrow \infty$,
where the heavy quark spin decouples from the light QCD degrees
of freedom, and the dynamics of the light quarks within a heavy
hadron is independent of the heavy quark flavor.
Consequently, heavy hadrons are organized in spin symmetry
multiplets, and flavor symmetry tells us how heavy hadron
observables scale with the heavy mass.
Corrections to the heavy quark limit are
treated systematically in the Heavy Quark Effective Theory (HQET) 
\cite{hqet1,hqet2,hqet3}.

For light quark flavors the Nambu-Jona--Lasinio (NJL) model
has been successfully used to describe the
dynamics of QCD. The properties of this model
are governed by global 
chiral invariance and its spontaneous breaking
in the ground state.
Indeed, employing appropriate functional integration
techniques the model can be reformulated as an
effective low-energy lagrangian in terms
of light 
pseudoscalar, vector and axial vector mesons fields,
which are described
surprisingly well,  embodying the
soft-pion theorems, vector dominance, Goldberger-Treiman and KSFR
relations and the integrated
chiral anomaly \cite{njl1}.
For a recent
review on these subjects see ref.~\cite{njl2} and references therein.

In this talk we shall review some recent work on
heavy meson physics based on an extension of
the NJL model \cite{extnjl}
which includes chiral symmetry for light quarks
and the heavy quark symmetries (for related work see also
\cite{extnjlrel}). 
In case of baryons, the diquark concept has proven
extremely useful to approximate the dynamics
at low energies. We will present our results
based on a phenomenological ansatz for an interaction
between heavy quarks and light diquarks \cite{toymodel}
and a Faddeev-equation involving composite light and
heavy diquarks \cite{faddeev}, 
respectively.

\section{Basic features of HQET}

In order to extract the heavy quark symmetries
from QCD one first identifies the relevant degrees
of freedom for the heavy quark field $Q(x)$. These
are given by the 'upper component'
\begin{eqnarray}
Q(x) &\to& Q_v(x) = \frac{1+\slash v}{2} \, e^{i\, m_Q \, \vx} \, Q(x)
\ .
\label{trafo}
\end{eqnarray}
Here the dependence on the heavy quark mass has been made
explicit by means of the heavy quark velocity $v_\mu$ with
$v^2=1$.
Upon integrating out the irrelevant degrees of freedom, the
(tree-level)
HQET Lagrangian can be obtained as a $1/m_Q$ series of
local operators \cite{hqet3}
\beq
\L{0}^{\rm HQET} &=& \bar Q_v \, (i \vD ) \, Q_v + O(1/m_Q)
\ ,
\label{lhqet}
\eeq
where $\D^\mu$ is the covariant derivative in QCD.
The operator $(i \vD)$ obviously does not depend on
the heavy quark spin and flavor, and therefore in the
limit $m_Q \to \infty$ we discover the heavy quark
spin and flavor symmetries.

One of the striking consequences of these symmetries can
be formulated as an analogue of the Wigner-Eckhart-theorem.
Using a matrix representation for the heavy pseudoscalar and
vector meson fields (describing the $B-B^*$ and $D-D^*$,
respectively\footnote{Here $\epsilon_\mu$ denotes the
polarization vector of the vector mesons with $v\!\cdot\!\epsilon=0$.})
$${\mathcal H}_v = \sqrt{M_H} \, 
\Pp \, \left[ i \gamma_5 + \slash \epsilon \right]$$
the hadronic matrix elements of any heavy quark current can
be expressed through
\beq
\langle \bar H(v) | \bar Q_{v'} \, \Gamma \, Q_v | H(v) \rangle
&=&
\xi(v\cdot v') \ \tr{\bar{\mathcal H}_{v'} \, \Gamma \, {\mathcal
H}_v}
\ .
\label{iw}
\eeq
Here the universal Isgur-Wise formfactor \cite{hqet2}
$\xi( v\cdot v') $ plays the role of a reduced
matrix element, and the Dirac trace can be viewed as
a Clebsch-Gordan coefficient. Furthermore for $v=v'$ the
normalization of the Isgur-Wise function is restricted
by the symmetries to be unity $\xi(1)=1$.
Similar statements are true for heavy baryons (see below).

\section{Heavy Mesons}

Let us consider an extended NJL model containing 
besides the free lagrangian of light  quarks ($q$)
and heavy  quarks ($Q_v$)
of definite velocity $v$,
 a four--quark interaction term which is motivated by the
general quark--current structure of QCD and obeys chiral
and heavy quark symmetries.
After Fierz--re\-arrange\-ment into the physical color--singlet
channel the relevant interaction term between a light quark 
and a heavy quark is given by 
(summation over repeated indices is understood) \cite{extnjl}
\begin{eqnarray}
  {\cal L}_{int}^{hl} &=&
      G_3
        \left(
      (\overline{Q}_vi\gamma_5 q)(\overline{q}i\gamma_5 Q_v) 
      -(\overline{Q}_v \gamma^\mu  q) P_{\mu\nu}^\perp 
        (\overline{q} \gamma^\nu Q_v)
        \right)
\nonumber \\[3mm]
&&   + G_3
        \left(
       (\overline{Q}_v q)(\overline{q} Q_v) 
      - (\overline{Q}_v  i\gamma^\mu \gamma_5 q) P_{\mu\nu}^\perp 
        (\overline{q}  i\gamma_5 \gamma^\nu Q_v)
        \right)
\ ,
\label{lint}  
\end{eqnarray}
where $P_{\mu\nu}^\perp = g_{\mu\nu} - v_\mu v_\nu$ is a
projection operator which is transversal with respect
to the heavy quark velocity, $v^\mu \, P_{\mu\nu} = 0$.

The bosonization procedure is standard \cite{njl1,njl2} and
consists in introducing composite mesonic fields $(\bar{q}q)$,
$(\bar{q}Q_v)$ into the generating functional which is
given by the path integral 
\begin{equation}
Z= \int {\cal D} q  {\cal D} \bar{q}
         {\cal D} Q_v  {\cal D} \overline{Q}_v 
e^{i \int d^4x (
{\cal L}_0 + {\cal L}_{int}^{ll}
  +{\cal L}_{int}^{hl} )}
\quad ,
\label{action}
\end{equation}
such that
the lagrangian becomes bilinear
in the quark fields and the latter can be integrated
out. 
Here ${\cal L}_{int}^{ll}$ is the usual interaction
term of light quarks. Symbolically, one has

\begin{center}
\psfig{file = 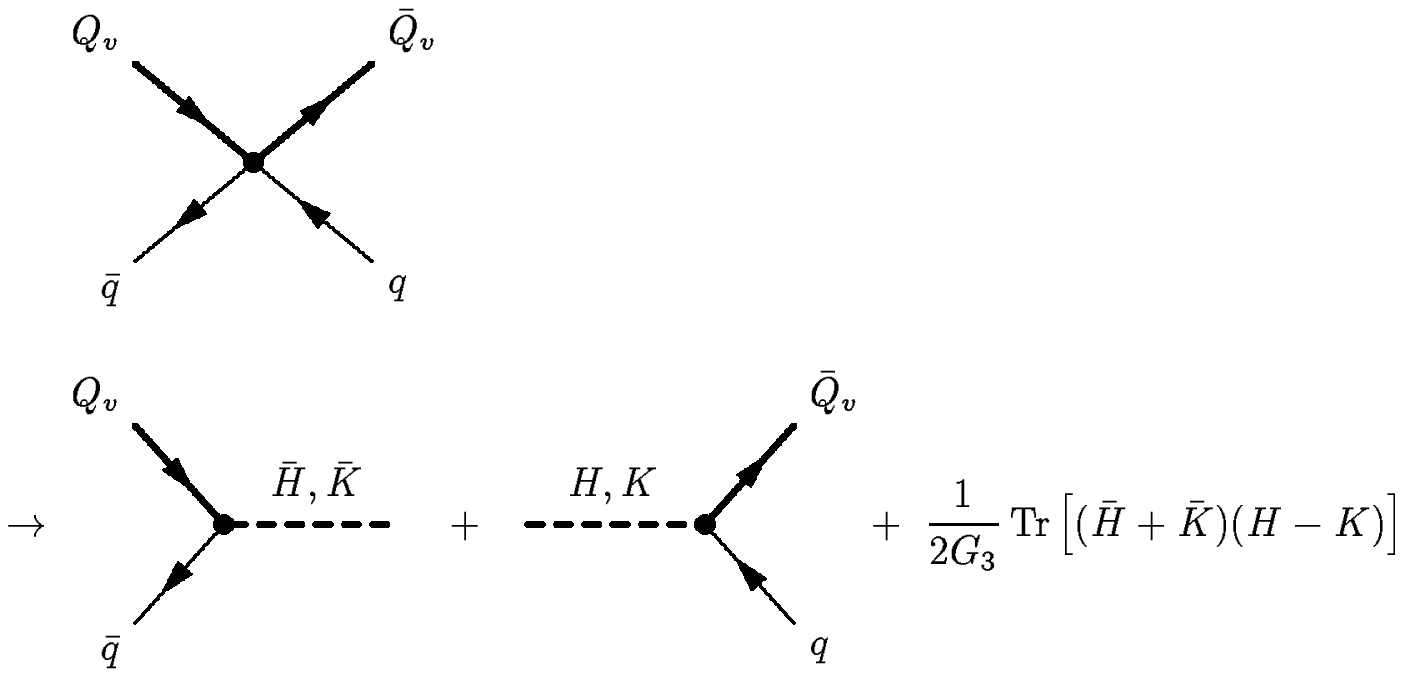 , bb = 100 550 520 760   , width = 10 cm}
\\
\end{center}

We use a non-linear representation where
$
  \xi = \exp (i \pi /F)
$
is an element in the coset space $SU(3)_L \times SU(3)_R$ $ /$
 $SU(3)_V$.
Here $F$ is the bare decay constant, and
$\pi=\pi^a \lambda^a_F/2$ represents the
light octet of (pseu\-do)\-Gold\-stone bosons
associated to spontaneous breakdown of chiral symmetry.
Our model includes light
vector mesons $V_\mu = V_\mu^a \lambda^a_F/2$
and axial-vector mesons $A_\mu = A_\mu^a \lambda^a_F/2$.

In the heavy-light sector we can collect
the pseudoscalar field $\Phi^5$ and the vector field $\Phi^\mu$
into a (super)field $H$ 
which represents the
$J^P=(0^-,1^-)$ doublet of spin symmetry. Analogously the
scalar field $\Phi$ and the axial-vector field $\Phi^{5 \, \mu}$
are combined in the (super)field $K$ describing the
$J^P=(0^+,1^+)$ doublet
\begin{eqnarray}
H &=& \Pp \, 
      (i \Phi^5 \gamma_5 + \Phi^\mu \gamma_\mu)
\quad , \quad v_\mu \Phi^\mu = 0 \quad ,\\
K &=& \Pp \,
      (\Phi  + i \Phi^{5\,\mu} \gamma_\mu \gamma_5)
\quad ,\quad v_\mu \Phi^{5 \, \mu} = 0 \quad .
\end{eqnarray}
Due to flavor symmetry
of HQET these fields describe both $B$ or $D$ mesons.
The details can be found in ref.\ \cite{extnjl}. 

Performing the integration over heavy and light quark fields 
in (\ref{action}) gives the quark determinant contributing
to the effective lagrangian as follows
\begin{eqnarray}
\L{eff} & = 
& - i N_c {\rm Tr} \ln i \slash{D} + \frac{1}{2G_3}
 {\rm Tr} \left[ (\overline{H} + \overline{K})(H- K) \right]
+ \ldots \quad ,
\label{det}
\end{eqnarray}
where 
\begin{eqnarray}
i \slash{D} &=& i \slash{\partial}
- \Sigma 
        + \slash{V} + \slash{A}\gamma_5 
        - (\overline{H}+\overline{K}) (i \vd )^{-1}
              (H+ K) 
\end{eqnarray}
is the quark Dirac operator containing light and heavy
meson fields.

To regularize the quark loops arising from 
(\ref{det}) we shall use a universal proper-time
cut-off $\Lambda$ which will be fixed from the light meson data.
For the applicability of the NJL-model to heavy
quark dynamics it is crucial to observe that the
typical relative momentum in a heavy hadron
is set by the light degrees
of freedom (gluons, light quarks) only, since all
reference to the heavy quark mass has been 
removed through \gl{trafo} and \gl{lhqet}.
Therefore it is the residual momentum of
the heavy quark $k_\mu = P_\mu - m_Q v_\mu$ which
is regularized by the cut-off $\Lambda$.

\subsection{Heavy meson self-energy and weak decay constant}


Expanding the term
$- i N_c {\rm Tr} \ln i \slash{D}$
in (\ref{det}) in powers of the meson
fields leads to the familiar loop expansion given by 
Feynman diagrams with
heavy and light mesons as external lines and heavy and
light quarks in internal loops.
For the light sector this has been done to derive an effective
lagrangian in terms of $\pi$, $\rho$ and $A_1$ fields
\cite{njl1}.
Comparison with experimental data fixes the parameters relevant
for the heavy sector:
the light constituent quark masses, $m^{u,d}= 300$ MeV,
$m^s = 510$ MeV, and a universal cut--off $\Lambda = 1.25$ GeV.

The self-energy part for the heavy mesons 
can be expanded 
in powers of the external momentum $v\cdot p$,
such that
the effective meson lagrangian acquires in configuration
space the desired form
\begin{eqnarray}
 {\cal L}_0^{H} & = & -  \, {\rm Tr} 
            \left[ \bar H \, (i \vd - \Delta M_H) \, H 
            \right]
    +  \, {\rm Tr} 
            \left[ \bar K \, (i \vd - \Delta M_K) \,  K 
             \right]
\quad ,
\end{eqnarray}
where the mass differences between heavy meson and heavy quarks are
$\Delta M_{H,K} = M_{H,K} - m_Q$.
Any left-handed weak decay current $\bar q_L \Gamma Q_v$ 
of a heavy meson $H$ 
can be represented by 
$$
J_\Gamma 
=
\frac{\sqrt{M_H} \, f_H}{2}
\,
{\rm Tr}\left[ \xi^\dagger \, \Gamma \, H \right]
$$
with $f_H$ being the weak decay constant defined
through
\begin{eqnarray}
    \langle 0| \bar{q} \gamma_\mu (1-\gamma_5) Q_v|H_v(0^-)\rangle 
  &=& i f_{H} M_{H} v_\mu \nonumber \quad .
\end{eqnarray}
In the extended NJL model it
can be expressed in a simple way by the $Z$-factors of
meson fields and
the four-quark coupling constant $G_3$
\begin{equation}
  f_{H} \sqrt{M_{H}} = \sqrt{Z_{H}} / G_3
\quad .
\label{law}
\end{equation}
We thus recover the familiar scaling of the weak decay constant
of heavy mesons with the heavy mass in HQET due to
heavy spin and flavor symmetry. A similar relation holds
also for the members of the $J^P=(0^+,1^+)$ doublet.
The experimental result for masses and decay constants
as a function of the four-quark coupling constant
$G_3$ is presented in Table~1. 

\begin{table}[hbt]
\caption{Table 1: Heavy meson parameters as a function of the coupling
constant $G_3$.
}
\begin{center}
\begin{tabular}{l|ccccc|c}
$G_3 [ \mbox{GeV}^{-2} ] $ &3&5&7&9&11& expected  \\
\hline
%
%
$\Delta M_H^s-\Delta M_H^{u,d} \left[ \mbox{MeV}\right]$ &
250 &160 & 110 & 80 & 50 & 100 \\
$ \Delta M_K^{u,d}-\Delta M_H^{u,d}\left[ \mbox{MeV}\right]$ &
370 & 390 & 410 & 430 & 450 & ? \\
%
%
$f_B\left[ \mbox{MeV}\right]$ &
300 &210 & 170 & 150 & 130 & 150--180 \\
$f_H^s/f_H^u$ &
1.13 & 1.13 & 1.12 & 1.11 &1.09 & 1.1--1.2 \\
\end{tabular}
\end{center}
\label{t1}
\end{table}

\subsection{Strong interactions with Goldstone bosons}

The Goldstone bosons $\pi$, $K$, $\eta$ couple to 
heavy meson fields via the vector and axial-vector
combinations
\begin{eqnarray*}
&& \V_\mu = \frac{i}{2} \, \left( \xi \, \partial_\mu \, \xi^\dagger
	+ \xi^\dagger \, \partial_\mu \, \xi \right)
\ ,
\quad
\A_\mu = \frac{i}{2} \, \left( \xi \, \partial_\mu \, \xi^\dagger
	- \xi^\dagger \, \partial_\mu \, \xi \right)
\ .
\end{eqnarray*}
The vector field $\V_\mu$ essentially couples with
a covariant derivative of $SU(3)_V$ 
$$
\partial_\mu \, H \to
D_\mu \, H =
\partial_\mu \, H
+ i \, H \, \V_\mu
$$
with small corrections due to the explicit breaking
$m_s \neq m_{u,d}$.
The coupling of the axial vector field 
$$
g_{HHA} \, {\rm tr}_D  \left[ \bar H \, H \, \slash{{\cal A}}^\pi \, 
	\gamma_5 \right]$$ 
is not restricted by the symmetries.
Our model predicts a value of about $g_{HHA} \approx - 0.2$
in accordance with experimental bounds
$\Gamma(D^*{}^+ \to D^0 \pi^+) < 0.131~\mbox{MeV}
\Rightarrow g_{HHA}^2 < 0.5$.

\subsection{Isgur-Wise function}

The calculation of the Isgur-Wise function defined
in eq.~\gl{iw} involves the following
Feynman diagram 
\begin{center}
\psfig{file = 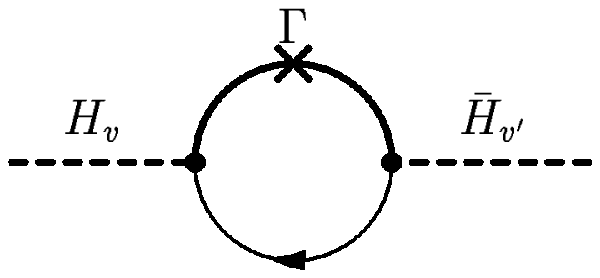 , bb = 86 680 260 760, width = 4 cm}
\end{center}
As a result we obtain for external momentum
$\vp = 0$ the general form
\beq
\xi(\omega) &=& a \, \frac{2}{1 + \omega} + (1-a) \, 
\frac{\ln(\omega +
\sqrt{\omega^2-1})}{\sqrt{\omega^2-1})}
\ , \quad  a = const. 
\eeq
Numerically for the slope parameter
of the Isgur-Wise function at the non-recoil
point $\rho^2 = - \xi'(1)$ we obtain 
the estimate $\rho = 0.67$. Other recent estimates
are slightly larger with $ 0.84 \leq \rho \leq 1.0$
(see \cite{Neubert} and references therein).

\section{Heavy baryons}

Light baryons have been recently described within
quark models as bound states of quarks and diquarks
\cite{njl2}. In ref.~\cite{toymodel} we have performed
a first step to obtain an analogous description of
heavy baryons (containing one heavy quark) in the
heavy quark limit by making use of the diquark picture.
Light diquarks are then coupled to the heavy quark
by a simple model interaction.
The lowest-lying baryons containing a single
heavy quark are classified according to their
flavor content (see Table~2).
The light quark quantum numbers (color, flavor, spin)
within a diquark are coupled as follows
\beq
&& 
3_C \otimes 3_C = \bar{3}_C \oplus 6_C
\ , \quad
3_F \otimes 3_F = \bar{3}_F \oplus 6_F
\ , \quad 
(1/2) \otimes (1/2) = 0 \oplus 1 \ .
\label{color}
\eeq

\begin{table}[hbt]
\caption{Table 2: Flavor content of the lowest-lying heavy baryons.}
\\
\begin{tabular}{l|l|l|cccc}
\hline
charmed baryon & spin partner &
bottom baryon & notation& $ij$ & I & S \\
\hline
$\Lambda_c^+ $ & trivial &
$\Lambda_b^0 $ & $T_v^{ij}$ &$\left[ud\right]$ & 0 & 0 \\
$\Xi_c^{+,0} $ & trivial  &
$\Xi_b^{0,-} $ &
$T_v^{ij}$ & $  \left[us\right]$, $\left[ds\right]$ & 1/2 & -1 \\
\hline
$\Sigma_c^{++,+,0} $ &
${\Sigma_c^*}^{++,+,0} 
$ &
${\Sigma_b^{(*)}}^{+,0,-} $ &
$S_v^{\mu\,ij}$ & $uu$, $\{ud\}$, $dd$ &
1 & 0 \\
${\Xi_c'}^{+,0} $ &
${\Xi_c^*}^{+,0}$ &
${\Xi_b^{'(*)}}^{0,-} $ &
$S_v^{\mu\,ij}$ &  $\{su\}$, $\{ds\}$ & 1/2 & -1 \\
$\Omega_c^0 $ & $ {\Omega_c^*}^0 $ &
${\Omega_b^{(*)}}^- $ &
$S_v^{\mu\, ij}$ &  $ss$ & 0 & -2 \\
\hline
\end{tabular}
\end{table}

Clearly, one has to consider only the color antitriplet
part in \gl{color} which together with the color triplet
of the heavy quark forms the physical color singlet baryon state.
The Pauli principle then requires the light
diquark system to be symmetric with
respect to simultaneous exchange of spin and flavor indices.
Therefore scalar diquarks 
only occur in the flavor antitriplet  $\bar 3_F$ and 
are hence realized
by an antisymmetric flavor matrix
$D^{ij}$, while axial-vector diquarks are realized by the
symmetric $6_F$ flavor matrix
$F^{ij}$.

Analogously, in the heavy mass limit we have
two types of heavy baryons:
In the first case
the two
  light quarks are coupled to spin zero, and
the heavy baryons are
  represented by an anti-symmetric flavor matrix of
  ordinary spin-1/2 Dirac spinors
\beq
\label{baryont}
T_v^{ij} &=&
\frac{i}{\sqrt{2}}
\left(
\begin{array}{ccc}
0     & \Lambda_Q & \Xi_{Q,\,I_3 = 1/2} \\
- \Lambda_Q & 0         & \Xi_{Q,\,I_3 = -1/2} \\
- \Xi_{Q,\,I_3 = 1/2} & - \Xi_{Q,\,I_3 = -1/2} & 0
\end{array}
\right) \ .
\eeq
      The two spin orientations of $T_v$ form
      the (in this case trivial) spin symmetry
      partners. \\
In the second case the two light quarks
are coupled to spin one and one obtains
a symmetric flavor matrix of
either spin-1/2 Dirac fields
      $B_v^{ij}$ or spin-3/2 Rarita--Schwinger fields
      ${B^*_v}^{\mu\,ij}$. These are then symmetry partners with
      respect to spin rotations of the heavy quark.
      Consequently, they can be combined in a multiplet
\beq
S_v^{\mu\,ij}& =& \frac{1}{\sqrt{3}} \gamma_5 (\gamma^\mu - v^\mu)
B_v^{ij} + {B^*_v}^{\mu\,ij} \ , \quad v_\mu S_v^\mu = 0
\no
\\
{B_v^{(*)}}^{ij}&=& \frac{1}{\sqrt{2}}\left(
\begin{array}{ccc}
\sqrt{2} \Sigma^{(*)}_{Q,\,I_3 = 1}
& \Sigma^{(*)}_{Q,\,I_3 = 0}
& \Xi^{'(*)}_{Q,\,I_3 = 1/2}\\
\Sigma^{(*)}_{Q,\,I_3 = 0}
& \sqrt{2} \Sigma^{(*)}_{Q,\,I_3 = -1}
& \Xi^{'(*)}_{Q,\,I_3 = -1/2} \\
\Xi^{'(*)}_{Q,\,I_3 = 1/2}
& \Xi^{'(*)}_{Q,\,I_3 = -1/2}
& \sqrt{2} \Omega_Q^{(*)} \ .
\end{array}
\right)
\eeq

\subsection{Heavy quarks coupled to light diquarks}

A straightforward phenomenological ansatz for
the interaction of heavy and light quarks in the
heavy baryon consists in treating the two light
quarks as elementary scalar and axial vector
diquark fields $D^{ij}$, $F_\mu^{ij}$, respectively. 
The interaction term which respects chiral and
heavy quark symmetries is expressed as \cite{toymodel}
\beq
\L{int} & = & \tilde G_1 \, \tr{\bar{Q}_v D^\dagger D Q_v} -
            \tilde G_2 \,  \tr{\bar{Q}_v F_\mu^\dagger
            \left(
		{}^{\left[1/2\right]}P^{\mu\nu} 
	       + {}^{\left[3/2\right]}P^{\mu\nu} 
            \right) F_\nu Q_v} \quad ,
\eeq
where $\tilde G_1$, $\tilde G_2$ are effective coupling
constant of dimension mass$^{(-1)}$ and 
$$
{}^{\left[1/2\right]}P^{\mu\nu}
= \frac{1}{3} \, (\gamma^\mu - \slash v v^\mu) (\gamma^\nu - \slash v
v^\nu)
\ , \quad
{}^{\left[3/2\right]}P^{\mu\nu}
= g^{\mu\nu} - v^\mu v^\nu - {}^{\left[1/2\right]}P^{\mu\nu} \ ,
$$
$$
v_\mu \, {}^{\left[1/2\right]}P^{\mu\nu} =
v_\mu \, {}^{\left[3/2\right]}P^{\mu\nu} =
\gamma_\mu \, {}^{\left[3/2\right]}P^{\mu\nu} = 0 \ .
$$
Again, we introduce the heavy baryon fields $T^{ij}$ and
$S_\mu^{ij}$ directly in the functional integral such that
the quark and diquark fields can be integrated out.
This leads to a quark-diquark loop expansion of the resulting
functional determinant (for details see \cite{toymodel}).
The self-energy part for the heavy baryon fields reads
\beq
\L{0}^{\rm eff} &=&
 {\rm tr_F} \, \bar T_v \, (i\vd - \Delta M_T) \, T_v
- \, {\rm tr_F} \, \bar S_v{}_\mu \, (i\vd - \Delta M_S) \,
S_v^\mu \ ,
\eeq
where the residual masses $\Delta M_{T(S)} = M_{T(S)} - m_Q$ are
related to the coupling parameters $\tilde G_1 \, (\tilde G_2)$,
the diquark masses $M_{D(F)}$ and the cut-off parameter used to
regularize the quark-diquark loops.

An interesting application of this simple model is the
calculation of the Isgur-Wise form factors for weak
heavy baryon transitions, which are defined by
the following expressions \cite{MRR}
\beq
&& A(v\cdot v') \;   \bar{T}_v  \Gamma T_{v'} \ ,
\qquad
 \left\{ B(v \cdot v') \, g_{\mu\nu} +
         C(v\cdot v') \, v_\mu' v_\nu
   \right\} \;  \bar{S}_v^\mu \Gamma S_{v'}^\nu \ .
\eeq
The Isgur-Wise formfactors $A(\omega)$ and $B(\omega)$ are
normalized at zero recoil $A(1) = B(1) = 1$.
Inserting an arbitrary weak current
into the heavy quark line and calculating the one-loop
Feynman diagrams,
we obtain
\beq
A (\omega) & = & r(\omega) =
\frac{\ln(\omega + \sqrt{\omega^2 - 1})}{\sqrt{\omega^2-1}} \ ,
\no \\
B(\omega) & = &
r(\omega) + \kappa\,
\frac{\omega -r(\omega)}{2+ \omega^2} \ ,
\qquad
C(\omega) = - \kappa \,
\frac{2 + \omega r(\omega)}{2 + \omega^2} \ ,
\eeq
with $\kappa > 0$ depending on the model parameters. 
Although the parameter range of our model cannot
be restricted sufficiently by experimental data
it is worth comparing our results with the
theoretical constraints imposed by Bjorken and Xu \cite{B92}.
Following the discussion in ref.\ \cite{KKP},
one obtains for $\omega \geq 1$
\beq
&&  A(\omega) \leq 1 \ ,\qquad
  A'(1) \leq 0 \ ,
\no \\
&& \frac{2}{3} \left|B(\omega)\right|^2 
+ \frac{1}{3} \left|\omega \, B(\omega) 
+ (\omega^2-1) \, C(\omega)\right|^2 \leq 1 \ ,
\label{BCconstraint1} \quad
B'(1)
\leq
 -\frac{1}{3} - \frac{2}{3} \, C(1) \ ,
\label{BCconstraint2}
\eeq
which is true in our model for any
positive value of $\kappa$.

\subsection{Light and heavy diquarks in the heavy baryon}

In a more sophisticated description of heavy baryons we
consider light {\bf and} heavy diquarks as {\bf composite}
fields, with their interaction given by the attractive 
diquark channel included in the extended NJL model (see
\cite{faddeev} for technical details).
Upon introducing composite diquark and baryon fields with
suitable constraints in the functional integral and performing
several functional integrations, we obtain a system of coupled
equations (Faddeev equation)
for the two configurations $ {\mathcal X} \sim
Q \, \left[ q \bar q^c \right]$ and $ {\mathcal Y} \sim
q \, \left[ Q \bar q^c \right]$ describing the heavy baryon
mass spectrum.
In our context the interaction takes place via quark exchange,
as it is illustrated below
\begin{center}
\psfig{file = 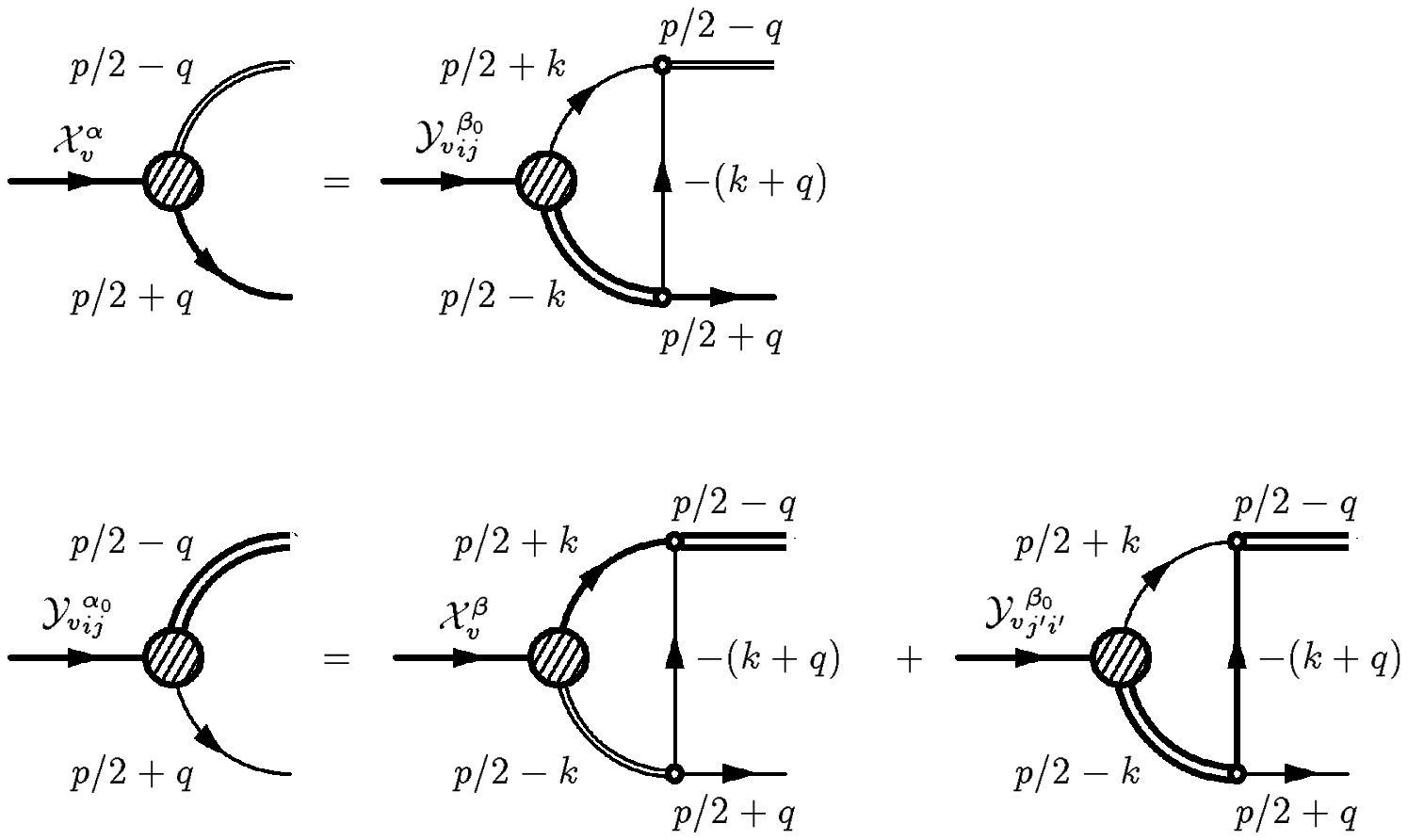 , bb = 100 450 550 710, width = 10 cm}
\end{center}

Using a static approximation to the quark exchange and 
fixing some model parameters from the light baryon sector,
we obtain a rough but reasonable estimate of heavy baryon masses
(see Table~3).
\begin{table}[htb]
\caption{Table 3: 
Heavy baryon mass splittings in MeV. 
The NJL coupling constant $G_1$,
which triggers the diquark properties, is given in units of the
coupling constant of the (light) pseudoscalar meson sector.}
\begin{center}
\begin{tabular}{l|llll}
$G_1/G_{\rm meson}$ & $ \Sigma_Q - \Lambda_Q $
& $\Xi_Q - \Lambda_Q $ & $ \Omega_Q - \Lambda_Q $ 
& $ \Xi^{'\,(\ast)}_Q - \Xi_Q $ \\
\hline
   1.20 &  160   &  190  &  490  & 135  \\
   1.30 &  175   &  200  &  515  &  145   \\
   1.40 &  190   &  200  &  530  &  155 \\
\hline
(Exp./Latt.) 
&  170 \cite{PD} & 185 \cite{PD}  
           &  455 \cite{WA62}  & 155  \cite{Latt} 
\\ 
\end{tabular}
\end{center}
\end{table}

\section{Conclusions}

In the present talk we have shown how chiral and
heavy quark symmetries can be implemented in 
relativistic quark and quark-diquark models in 
order to describe the properties of heavy hadrons.
We have investigated the extended NJL model, including
$SU(3)_F$ breaking and discussed the resulting 
mass spectrum for heavy mesons of the spin/parity
doublets $J^P=(0^-,1^-)$ and $J^P=(0^+,1^+)$, their
strong coupling constants and weak decay constants
and the Isgur-Wise form factors.

Heavy baryons can be reasonably described within
a model where the heavy quarks couple to elementary
light scalar and axial vector diquark fields. 
The predicted form of the Isgur-Wise functions
obeys the Bjorken/Xu inequalities in a parameter-independent
way.
On the basis of composite light and heavy diquark fields
within the heavy baryon, a Faddeev equation is derived from
the extended NJL model, which gives a reasonable description
of the heavy baryon spectrum. 

In order to improve the results on interactions, decays,
Isgur-Wise functions etc., their remain several tasks to
be done.
One important question is how to settle the lack of confinement
in NJL-type models which influences the external momentum-dependence
of hadron properties. Another interesting problem is to what extent
$1/m_Q$ corrections can be studied in such a framework.

We would like to express our gratitude to our co-authors
H.~Reinhardt, R.~Friedrich and C.~Kettner for their
contribution to these investigations and fruitful
cooperation.

\end{document}